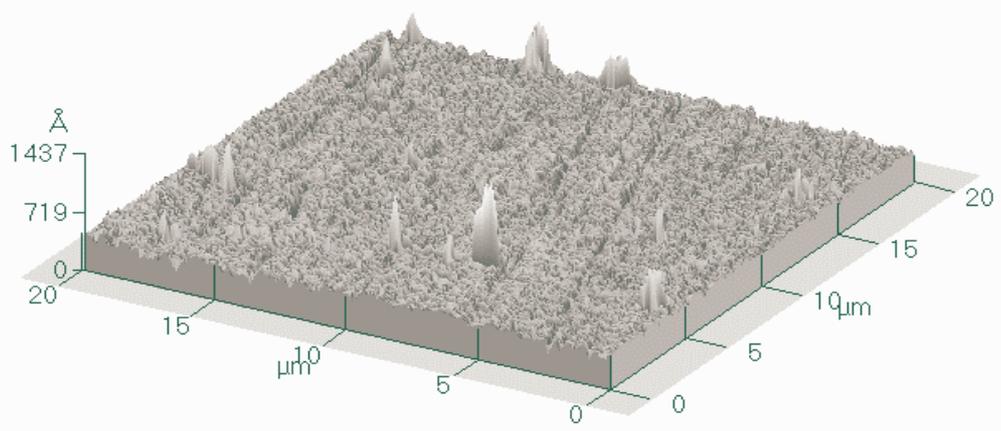

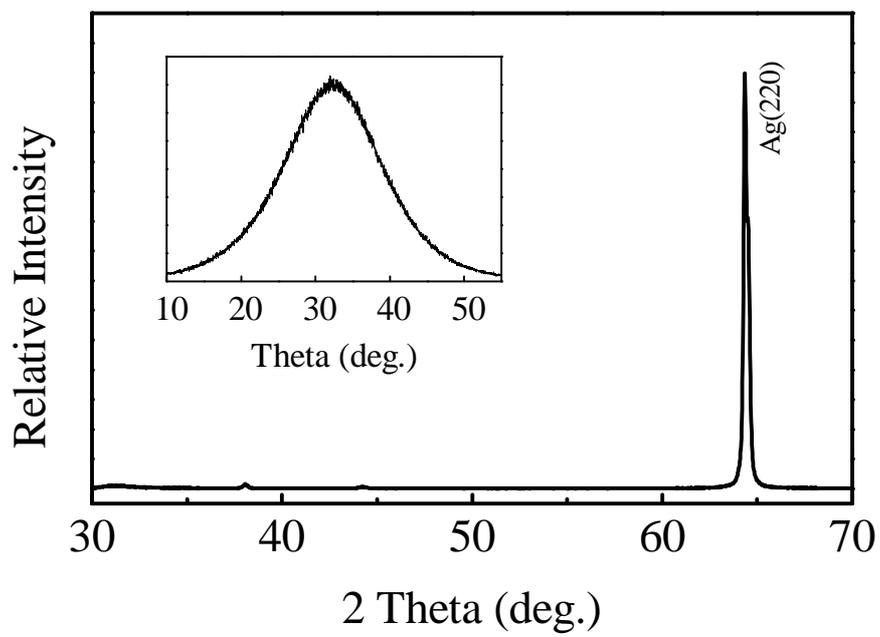

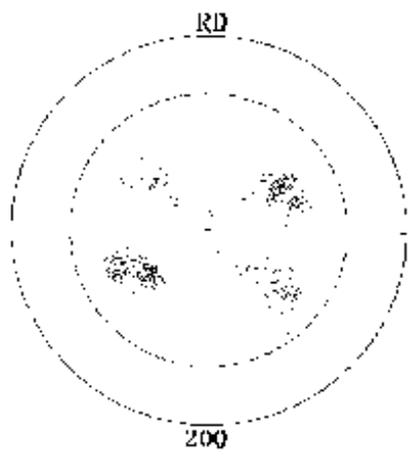

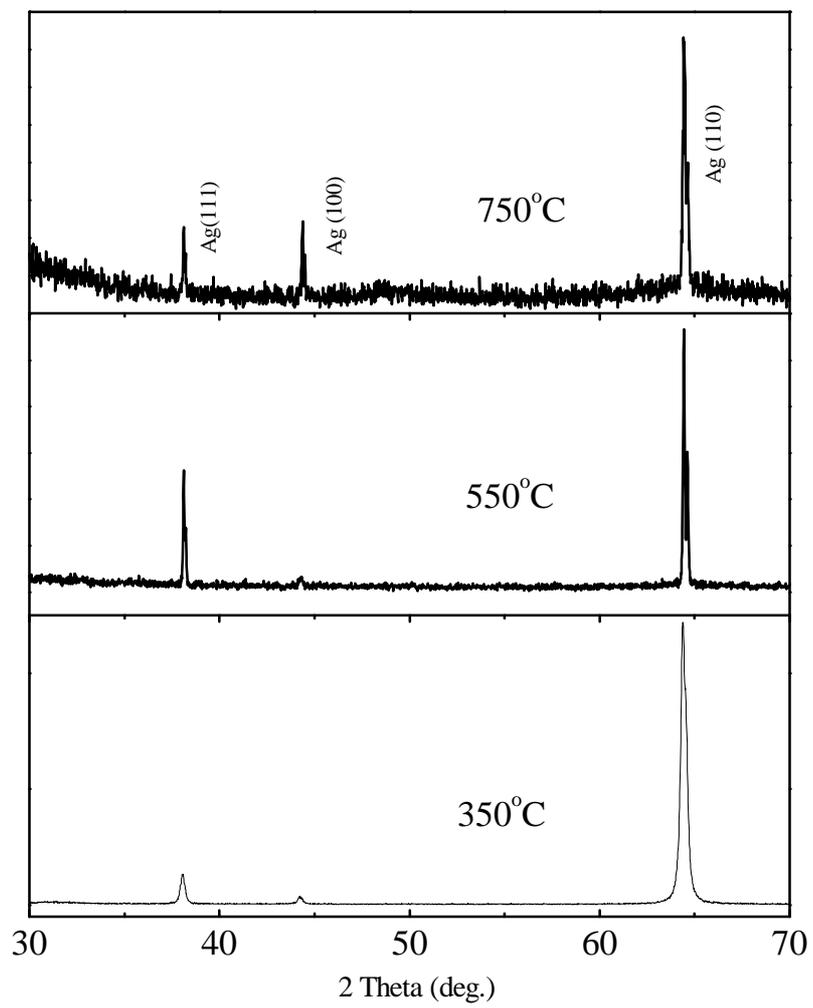

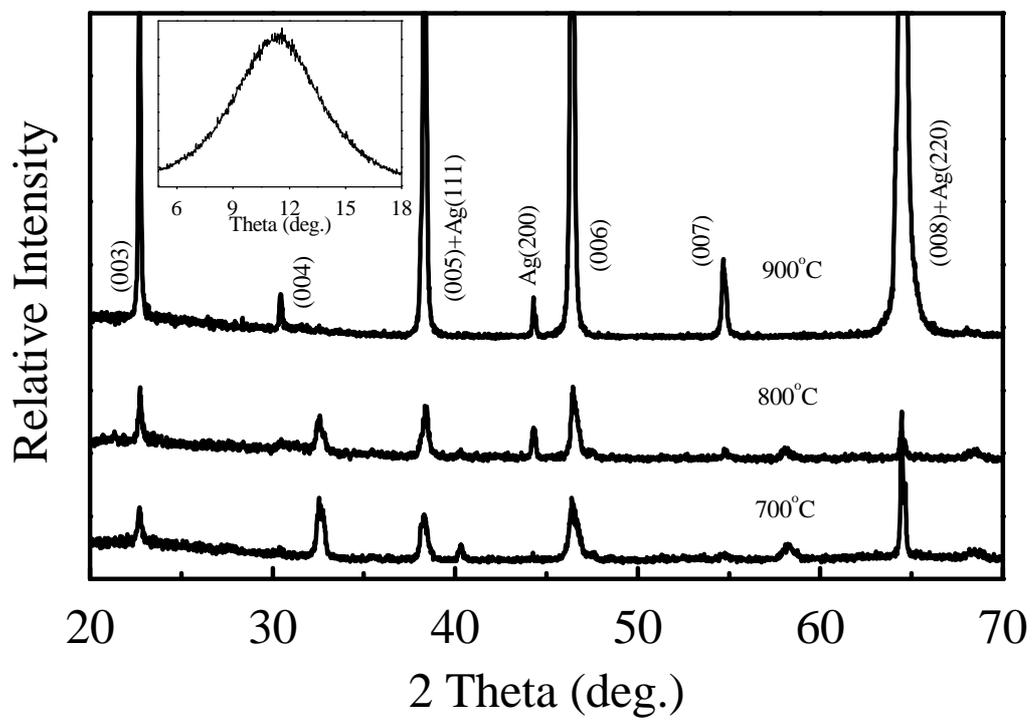

Figure 3(a). Rong-ping Wang *et al*.

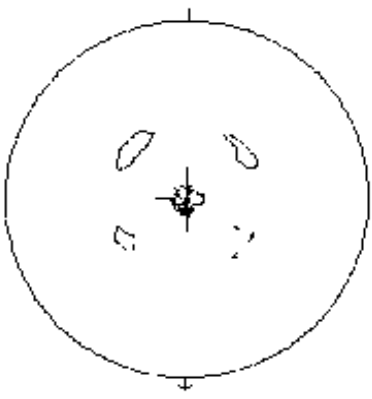

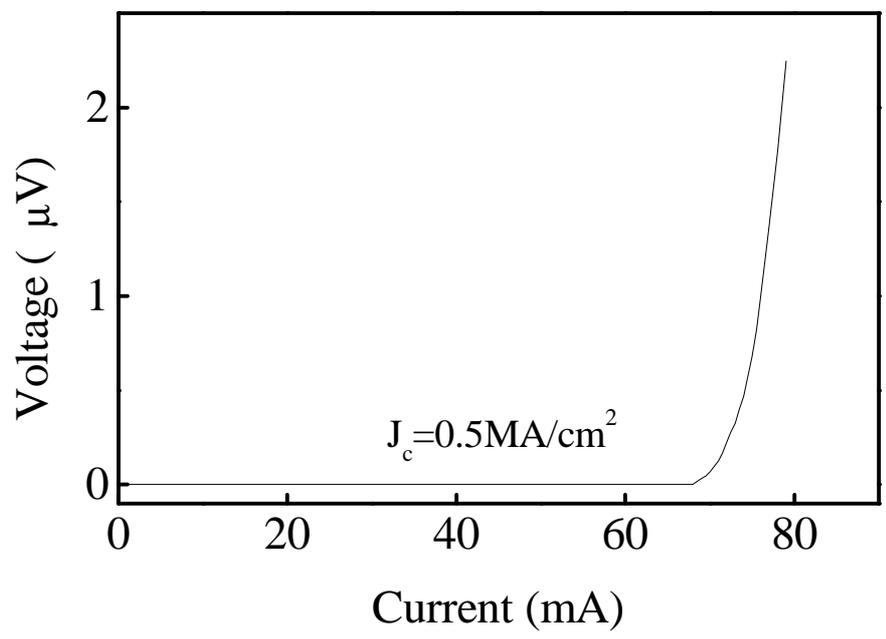

# A New Direct Process to Prepare $YBa_2Cu_3O_{7-\delta}$ films on Biaxially Textured Ag{110}<211>

Rongping Wang[a)], Yueliang Zhou, Shaohua Pan, Meng He, Zhenghao Chen and Guozhen Yang

*Laboratory of Optical Physics, Institute of Physics & Center for Condensed Matter Physics, Chinese Academy of Sciences, Beijing 100080,P.R.China*

Abstract

YBCO films were successfully prepared on biaxially textured Ag{110}<211> substrates by using pulsed laser deposition. X-ray diffraction results showed that the degree of preferential orientation of Ag{110}<211> substrates varied with increasing annealing temperature. With a thin template layer deposited at low temperature, YBCO film with c-axis orientation and in-plane biaxial alignment could be obtained at high deposition temperature. Scanning electron microscopy observation revealed that YBCO grains enlarged but Ag grains on the surface of the YBCO films became smaller with increasing deposition temperature. At optimal deposition conditions, Ag atoms diffuse into the YBCO grain boundaries, and then fill in the weak-link regions in the YBCO film, resulting in the conduction easier. $J_c$ value of $5\times10^5 A/cm^2$ was obtained at 77K and zero magnetic field for the best YBCO film in our work.

a). Electronic mail: xyw@aphy.iphy.ac.cn

# 1. Introduction

The large scale applications of high temperature superconducting (HTSC) materials such as wires for power transmission cables and solenoids are clearly limited by its ceramic nature. Deposition on metal substrates with or without buffer layers has been proved to be one of the best approaches to overcome these problems in HTSC materials[1,2]. Using an auxiliary ion gun, Wu *et al.* [3] prepared highly textured YSZ buffer layers on Ni substrates, and then deposited 2-μm thick $YBa_2Cu_3O_{7-\delta}$ (YBCO) films with a critical current ($J_c$) of $10^6 A/cm^2$ at 77K. Also, D.P.Norton *et al* [4] prepared high-$J_c$ YBCO tapes by epitaxial deposition on rolling-assisted biaxially textured substrates with a $J_c$ in the order of $10^6 A/cm^2$ at 77K. These novel approaches have shown potential for developing long conductors for industrial applications. However, the Ni alloy which they used as a substrate is easily chemically reactive with respect to YBCO, resulting in the degradation of the superconducting properties, so that one or more buffer layers are necessary in order to prevent interdiffusion. Unfortunately, these buffer layers complicate the preparation process and limit large-scale production. Hence, a more simple approach to prepare high-$J_c$ YBCO films is desirable.

Previous work showed that Ag is inert with respect to YBCO. By using Ag as substrates, YBCO thin films have been grown by various methods[5-8]. In the most cases, $J_c$ values were no more than $10^4 A/cm^2$, which was too low for large-scale applications. The main reasons for lower $J_c$ values in YBCO films appear to be as follows: one is the growth methods, the YBCO films were obtained by using a non-vapor approach in references[5-8], where the optimal oxygen content in YBCO films was obtained by post-annealing at high temperature and ambient oxygen. In general, the superconducting properties for YBCO films prepared by non-vapor deposition and post-annealing procedure are poorer than those obtained by vapor deposition and *in-situ* annealing procedure. Another reason is that the HTSC films have a so-called weak-link effect at grain boundaries, which can severely limit their applications. For YBCO, c-axis texture alone is not sufficient enough to overcome the weak-link problem, an in-plane biaxial alignment is required for YBCO films with high-$J_c$. However, to our knowledge, no any data about biaxial alignment in these works were presented.

It has been proved that pulsed laser deposition (PLD) is one of the best deposition methods for preparing HTSC films[9]. As a vapor deposition approach, it can effectively avoid the oxygen deficiency in HTSC films. Using PLD, M.Yamazaki *et al.* prepared YBCO superconducting tapes on textured Ag(110) substrates with a $J_c$ of $1.2 \times 10^5 A/cm^2$ at 77K, which was the best reported results to our knowledge[10]. In this paper, we report that YBCO films with a $J_c$ of $5 \times 10^5 A/cm^2$ were successfully prepared on biaxially textured Ag{110}<211> substrates by PLD. X-ray diffraction (XRD) results indicated that the degree of preferential orientation of Ag{110}<211> substrates varied with increasing annealing temperature, and that YBCO films with c-axis orientation and in-plane biaxial alignment could be obtained at suitable deposition conditions. Scanning electron microscopy (SEM) observation revealed that YBCO grains enlarged but Ag grains on the surface of the YBCO films became smaller with increasing deposition temperature. The effect of the deposition temperature on the characteristics of the YBCO films including texture and microstructure was analyzed.

**2. Experiments**

Biaxially textured Ag{110}<211> substrates were prepared by cold-rolling 99.99% pure silver followed by recrystallization at 700°C for 0.5 hours. The thickness of Ag{110}<211> substrates used was 200μm. Substrates were rinsed in organic solvents and de-ionized water before being put into deposition chamber. YBCO films were deposited by PLD. The target was a 30mm-diameter sintered pellet of high purity YBCO. A round Ag foil with the same diameter of the YBCO target was cut into fan-shaped foil with a central angle of 30°. Then the fan-shaped Ag foil was attached on the surface of the YBCO target. A 308nm XeCl excimer laser was focused into the vacuum chamber to an energy density of 2.5 to $3.0 J/cm^2$ at the target with a repetition rate of 3Hz. The distance from the target to substrate was 3.5cm. The substrate temperature was measured by infrared pyrometer. During all deposition processes, the deposition chamber was prepumped to $2\times10^{-3}$Pa, and then introduced $O_2$ to 50Pa pressure. The YBCO films with the thickness of 1μm were deposited at temperatures from 650°C to 920°C. A standard XRD θ-2θ technique was employed for the crystallographic phase analysis and detection of the preferential orientation of the deposited films. By using Cu $K_\alpha$ radiation, diffracted intensity and the full width at half-maximum (FWHM) were recorded at 0.02° intervals over a range of 20°--70°. The X-ray pole figure was used to determine the in-plane biaxial alignment. SEM was employed to observe the surface morphology. The superconducting properties of the as-grown films were measured by a standard four probe method.

### 3. Results and Discussion

Figure 1(a) presents the surface morphology of Ag{110}<211> substrate observed by atomic force microscopy. Some convex grains and ravines can be seen. Average line scans in a 20μm×20μm region indicate a root-mean-square roughness of 14.7nm, which is a considerably flat surface for an unpolished metallic substrate. XRD result for Ag{110}<211> substrate is shown in Fig.1(b), where only intense Ag(220) peak appears while Ag(111) and Ag(200) peaks are almost unseen, indicating that Ag{110}<211> substrate has good out-of-plane orientation. This is also confirmed by the inset of Fig.1(b), where the FWHM of rocking curve of Ag(220) peak is 14.5°. Fig.1(c) is (200) pole figure of Ag{110}<211> substrate. In general, only two strong diffraction spots should appear in the standard (200) pole figure of Ag{110}<211> substrate. However, it is difficult to avoid some twin related components in the rolling and recrystallization texture in Ag{110}<211> substrate[11]. In Fig.1(c), the other two weak diffraction spots come from the twin components in Ag{110}<211> substrate. The azimuthal distributions evaluated from the FWHM of (200) poles are from 15° to 21°. Such Ag{110}<211> substrates with good in-plane biaxial alignment offer a solid background for preparing high-$J_c$ YBCO films.

A.Goyal *et al*. [11] have reported that the texture of Ag substrates is unstable at high temperature which is suitable to deposit YBCO films. To confirm this point,

Ag{110}<211> substrate was put into a vacuum chamber with $3\times10^{-3}$Pa pressure and was annealed at various temperatures for 0.5h. Then it was slowly cooled to room temperature, and finally the orientation of the substrate was detected by XRD as shown in Fig.2. It can be seen that with increasing annealing temperature, Ag(111) and (200) peaks appear in sequence, however, the substrate was (220)-preferential orientation under the annealing temperature and time we used. It can be expected that the intensities of Ag(111) and (200) peaks increases with increasing annealing temperature and time. Moreover, the appearance of Ag(111) and (200) peaks indicates that the in-plane biaxial alignment of Ag{110}<211> substrate is easily destroyed at high temperature.

It has been proved that YBCO films can be grown epitaxially on (220), (111) or (200)-oriented Ag single crystal substrates[12]. If we deposit YBCO films on Ag{110}<211> substrates at 700°C--800°C, it is difficult to obtain YBCO films with good in-plane biaxial alignment due to the texture unstability of Ag{110}<211> substrates at high temperature as shown in Fig.2. That is, although highly c-axis oriented YBCO films can be obtained, a poor in-plane biaxial alignment for these films will appear, resulting in weak link in YBCO films and hence degrading superconducting properties. On the other hand, if we deposit YBCO films at low deposition temperature, high quality YBCO films are difficult to obtain due to poor crystallinity at low temperature. To solve this dilemma is a key to directly prepare YBCO films on Ag{110}<211> substrates.

We prepared YBCO films according to the following procedure: First, a thin YBCO layer (10nm) was deposited on Ag{110}<211> substrate at 500°C and 50Pa oxygen pressure, then the deposition temperature was raised to higher temperatures. YBCO films with a thickness of 1μm were deposited at these temperatures. Due to the low deposition temperature at the early stage, the variation of the texture of Ag{110}<211> substrates was smaller. XRD patterns showed that the thin YBCO layer was crystalline state with (00*l*) orientation. It was expected that the thin YBCO layer could be used as a good template layer for depositing YBCO films at high temperatures. Figure 3(a) shows the typical XRD patterns for YBCO films deposited at temperatures from 700°C to 900°C. At lower deposition temperature, YBCO film had (00*l*) preferential orientation, and the YBCO (103) peak also existed. When the temperature was raised to 900°C, YBCO film had almost complete (00*l*) orientation. A comparison of the intensities of (00*l*) peaks of YBCO films deposited at high and low temperature indicates that high temperature is beneficial to form the c-axis oriented YBCO film. As mentioned earlier[6], Ag atoms easily diffuse into YBCO films at high temperature, and the amount of the metal is larger in the neighbourhood of the substrate and decreases exponentially along the normal of the interface. In this work, Ag content at the neighbourhood of the interface is almost the same as that at the surface of the YBCO film due to the rotatable target attached with a fan-shaped Ag foil. The identically distributed Ag content lowers the partial melt (peritectic) temperature of Y-123 or reduces the recrystallization temperature of Y-123, causing preferential nucleation and subsequent growth in the c-axis oriented film, hence the optimal deposition temperature (900°C) for our c-axis oriented YBCO film as shown in Fig.3(a) is somewhat lower than those reported values[5-8]. The inset of Fig.3(a) shows the rocking curve of the YBCO (003) peak, the FWHM is 4.5°. In comparison with the FWHM of the rocking curve of Ag(220) peak, the improved out-of plane orientation is evident. Fig.3(b) shows the pole figure of (103) planes of YBCO film deposited on Ag{110}<211> substrates at 900°C. Obviously, four diffraction spots are symmetric but their intensities are not uniform. The main reason may be due to smaller size of the sample and the existence of the in-plane misalignment YBCO grains. However, these four diffraction spots hint that in-plane biaxial alignment YBCO film can be obtained by further optimized deposition conditions.

Figures 4(a),(b) and (c) present the surface morphology of YBCO films deposited at 700°C,800°C and 900°C, respectively. At low deposition temperature such as 700°C, a large number of small round grains appeared in the SEM photograph, and instinct grain boundaries existed. Energy dispersive X-ray analysis indicated that the main composition of these round grains was Ag. When the temperature was raised to 800°C, the YBCO grains became larger while the Ag grains prolonged and tended to fill in the grain boundary between the YBCO grains. The square shape YBCO grains appeared at 900°C,

and an obvious layer-like morphology was evident from Fig.4(c), moreover, Ag grains almost disappeared. We infer that Ag grains become smaller at high temperature and diffuse into the YBCO grain boundaries, and fill in the weak-link regions in the YBCO film, resulting in the easier conduction.

The transport properties of YBCO films were measured by the standard four-probe method in 77K and zero magnetic field using a 1μV/cm criterion. Fig.5 shows the voltage vs. current plot for a typical YBCO sample deposited at 900°C and 50Pa oxygen pressure. The sample size was 15mm×2mm with a patterned bridge 100μm long and 30μm wide. The distance of probes was 3mm. A respectable $J_c$ values could be obtained up to $5×10^5 A/cm^2$, which was comparable to some of the reported data on YBCO thin films on ceramic substrates. We expect the $J_c$ value to increase significantly as we further optimize deposition conditions.

## 4. Conclusions

In summary, YBCO films were successfully prepared on biaxially textured Ag{110}<211> substrates. XRD results showed that the degree of preferential orientation of Ag{110}<211> substrates varied with increasing annealing temperature. With a thin template layer deposited at low temperature, the YBCO film with c-axis orientation and in-plane biaxial alignment could be obtained at high deposition temperature. SEM observation revealed the surface morphology evolution of YBCO films deposited at three typical temperatures. It was found that YBCO grains enlarged but Ag grains on the surface of the YBCO films became smaller with increasing deposition temperature. Under optimal deposition conditions, Ag atoms diffuse into the YBCO grain boundaries and improve the weak-link in YBCO film. A respectable $J_c$ values of $5×10^5 A/cm^2$ measured by standard four-probe method in 77K and zero magnetic field for the best YBCO film was obtained in our work.

**Acknowledgments**

This work was supported by the National Center for Research and Development on Superconductivity of China.

**Captions**

Figure 1: Ag{110}<211> substrate. (a) AFM photograph in the range of 20×20μm$^2$, (b) X-ray diffraction pattern, the inset shows the rocking curve of Ag(220) peak,(c) the (200) pole figure.

Figure 2: X-ray diffraction patterns of Ag{110}<211> substrate annealing at (a) 350°C, (b)550°C, (c)750°C.

Figure 3: (a) X-ray diffraction patterns of YBCO films deposited at the typical temperatures, the inset shows the rocking curve of YBCO(003) peak, (b) the (103) pole figure of YBCO film deposited at 900°C and 50Pa oxygen pressure.

Figure 4: SEM micrographs of YBCO films deposited at the typical temperatures (a) 700°C, (b) 800°C, (c) 900°C.

Figure 5: Plot of the voltage vs. current for a typical YBCO film deposited at 900°C and 50Pa oxygen pressure.